\documentclass{article}

\makeatletter 
\@addtoreset{equation}{section}
\makeatother  

\usepackage{indentfirst}
\usepackage{setspace}
\usepackage{amsmath, mathtools}
\usepackage{mathrsfs}
\usepackage[colorlinks, linkcolor=black, anchorcolor=black, citecolor=black]{hyperref}
\usepackage{graphicx}
\usepackage{tocloft}
\title{\textbf{\huge{{Thermodynamics of Schwarzschild-Beltrami-de Sitter Black Hole}}}}
\author{Hang Liu$^{a}$\thanks{E-mails:hangliu@mail.nankai.edu.cn} and Xin-he Meng$^{a,b}$\thanks{E-mails:xhm@nankai.edu.cn}
\\
\\
$^{a}$ \itshape School of Physics, Nankai University, Tianjin 300071, China\\
$^{b}$ \itshape State Key Laboratory of Theoretical Physics, Institute of Theoretical Physics,\\
\itshape Chinese Academy Of Science, Beijing 100190,China}
\date{}
\begin{document}
\large
\maketitle

\begin{abstract}
In this paper, we investigate the thermodynamical properties of Schwarzschild-Beltrami-de Sitter black hole introduced by Mu-Lin Yan \textit{et al.} in 2013 by introducing inertial Beltrami coordinates to traditional non-inertial Schwarzschild-de Sitter metric which is the exact static spherical symmetry solution of Einstein equation with a positive cosmological constant $\Lambda$. Based on this new metric, we compute entropy on all horizons and we give the entropy bound of the black hole. Hawking temperatures are calculated  by considering a perturbation to entropy relations due to that the spacetime described by this inertial coordinates is no longer a stationary spacetime in which surface gravity related to Hawking temperature is defined well on killing horizon. We also get the Smarr relations and the first law of thermodynamics. We find that the Schwarzschild-Beltrami-de Sitter black hole seems to have similar thermodynamical properties to Schwarzschild-de Sitter black hole in the comparison between their corresponding thermodynamical quantities, although the new black hole metric is described by inertial coordinates which exclude the effects of inertial force.
\end{abstract}

\setcounter{page}{0}
\tableofcontents\thispagestyle{empty}
\newpage
\section{Introduction}
Recently, Mu-Lin Yan \textit{et al.} give a dynamical black hole solution \cite{mulin}, Schwarzschild-Beltrami-de Sitter(S-BdS) metric, which is the solution of Einstein filed equation with a positive cosmology constant $\Lambda$ obtained by introducing the inertial Beltrami coordinates to the existed Schwarzschild-dS(S-dS) black hole metric under the consideration of coordinates condition. In the spacetime described by inertial Beitrami metric, the motion of free particle is inertial ,i.e. the coordinate velocity of a geodesic is constant when mass $\mathcal M\to 0$ since there is no gravity source exists. This is the crucial differences between Schwarzschild-Beltrami-de Sitter metric and Schwarzschild-de Sitter metric which is non-inertial implying that inertial motion law for free particle does not hold even though the gravity source $\mathcal M$ disappears in the S-dS spacetime. This fact can be interpreted as the inertial Beltrami metric describes the spacetime without fake inertial forces, i.e. Schwarzschild-Beltrami-de Sitter metric describes the effects of pure gravity without inertial forces while the old Schwarzschild-de Sitter metric describes the mixing effects of gravity and inertial forces \cite{ams}, which is also the main motivation of introducing inertial Beltrami coordinates.

In  Minkowski spacetime, the vacuum defined in an inertial coordinate is extremely different from the vacuum defined in a uniformly accelerating coordinate, and the amazing difference is that an uniformly accelerating observer in Minkowski spacetime can detect thermal radiation while the inertial observer cannot. This is the famous Unruh effect proposed and proved by physicist William Unruh et al. based on Hawking Radiation theory and equivalence principle. Actually, the spactime diagram of Rindler spacetime(i.e. Minkowski spacetime covered by accelerating coordinate system ) is highly similar to Schwarzschild spacetime described by Kruskal coordinate and this may reflect the local equivalence between gravity and acceleration. The root of the Unruh effect, which is regarded as a kind of quantum effect, is that the time $\eta$ defined by Rindler observer is different from time $t$ defined by inertial observer. By quantum field theory, we know that both creation and annihilation operator related to the definition of vacuum state are dependent of time implying that different time coordinate corresponds to different vacuum state. What's more, the zero point energy of vacuum state defined by time $\eta$ of Rindler observer is lower than that defined by time t of inertial observer, and the difference of this two zero point  energy will lead to observable effect which makes the Rindler observer feel a “thermal bath” around.
In paper \cite{Guo1}, the authors argued that ``there are special coordinate systems in dS-spacetime, called the Beltrami coordinate systems, which play the role of Minkowski coordinate systems in Minkowski spacetime and argue that the horizon in the dS-spacetime with a Beltrami metric should be at zero temperature and that the relation between Beltrami and static metrics of dS-spacetime is analogous to that between Minkowski and Rindler metrics of flat spacetime.'' Inspired by this perspective, we think that there may also exist Unruh-like effect in dS and S-dS spacetime, the zero point energy of vacuum defined by Beltrami inertial observers is different from that defined by non-inertial observers in S-dS static spactime as the situation in Minkowski flat spacetime we discussed above. Under this consideration, it is clear to see that when considering the thermodynamics of S-dS black holes, we actually do a calculation under the mixing effect of gravity and the inertial force corresponding to Unruh effect similarly rather than pure gravity effect.

The simplest spherical symmetry vacuum solution of Einstein's field equations with a positive cosmology constant was derived by Kottler(1918), Weyl(1919), Trefftz(1922), which is called Schwarzschild-de Sitter metric with the line element
\begin{equation}
ds^2=(1-\frac{2\mathcal M}{r}-\frac{r^2}{\ell^2})dt^2-(1-\frac{2\mathcal M}{r}-\frac{r^2}{\ell^2})^{-1}dr^2-r^2(d\theta^2+\sin^2(\theta)d\phi^2)
\end{equation}
In order to make this non-inertial metric reformed into an inertial metric which allows inertial motion of free particle when $\mathcal M\rightarrow 0$, Mu-Lin Yan \textit{et al.}  resolve the vacuum Einstein equation of dS-GR with the requirement of gravity-free metric \cite{mulin} and finally get the S-BdS metric written in the inertial Beltrami coordinates and this is a new form of S-dS metric. Its physical meaning and possible applications are discussed in Ref.\cite{mulin}. The coordinates transformation between the inertial Beltrami coordinates and the  non-inertial coordinates is
\begin{equation}
\begin{aligned}\label{1}
r_N&=\frac{r}{\sqrt{1+\frac{r^2-t^2}{\ell^2}}}\\
t_N&=\int\frac{dt}{1-\frac{t^2}{\ell^2}}=\ell\arctan{\frac{t}{\ell}}\\
\theta_N&=\theta\\
\phi_N&=\phi
\end{aligned}
\end{equation}
coordinates$(t,r,\theta,\phi)$ and $(t_N,r_N,\theta_N,\phi_N)$ represent the new inertial coordinates and the old non-inertial coordinates, respectively. The parameter $\ell$ stands for cosmological radius which is related to cosmological constant by $\Lambda=3\ell^{-2}$.
\\
By employing (\ref{1}), we can get the metric transformation
\begin{equation}
g_{\mu\nu}(x)=\frac{\partial x_N^\alpha}{\partial x^\mu}\frac{\partial x_N^\beta}{\partial x^\nu}g_{\alpha\beta}(x_N)
\end{equation}
which finally yields  the Schwarzschild-Beltrami-de Sitter metric \cite{mulin}
\begin{equation}
\begin{split}
ds^2&=g_{\mu\nu}(x)dx^\mu dx^\nu\\
&=\left(\frac{1-\frac{r^2}{\ell^2\sigma}-\frac{2\mathcal M\sqrt{\sigma}}{r}}{(1-\frac{t^2}{\ell^2})^2}-\frac{\frac{r^2t^2}{\ell^4}}{(1-\frac{r^2}{\ell^2\sigma}
-\frac{2\mathcal M\sqrt\sigma}{r})\sigma^3}\right)dt^2\\
&\quad-2\frac{\frac{rt}{\ell^2}(1-\frac{t^2}{\ell^2})}{(1-\frac{r^2}{\ell^2\sigma}-\frac{2\mathcal M\sqrt\sigma}{r})\sigma^3}dtdr
-\frac{(1-\frac{t^2}{\ell^2})^2}{(1-\frac{r^2}{\ell^2\sigma}-\frac{2\mathcal M\sqrt\sigma}{r})\sigma^3}dr^2\\
&\quad-\frac{r^2}{\sigma}(d\theta^2+\sin^2\theta d\phi^2)\label{metric}
\end{split}
\end{equation}
\\
where the $\sigma$ is defined to be

\begin{equation}
\sigma=1+\frac{r^2-t^2}{\ell^2}
\end{equation}
\\
One can check that this new metric satisfies the Einstein field equations in empty spacetime(cosmological constant plays the role of geometry terms putted on the left side of Einstein equation). It is also easy to see that when $\ell\rightarrow\infty$, it coincides with the Schwarzschild metric and when $\ell\rightarrow\infty$, $\mathcal M\rightarrow0$, it would go back to Minkowski metric in a spherical coordinate system, and for $\mathcal M\rightarrow 0$ we will obtain BdS metric. One may note that the cost of introducing inertial S-BdS metric describing pure gravity effects is that this new metric is no longer static, not even stationary since the metric $g_{\mu\nu}$ is time dependent and $g_{tr}\neq0$ which indicates that the S-BdS black hole is a dynamical black hole rather than static.

To have a better understanding of S-BdS black hole, would like to give a brief review of de Sitter Special Relativity (dS-SR) by reviewing Ref.\cite{yan2,mulin,Guo1,Guo2,Guo3}. The form of the Lagrangian for a free particle is
\begin{equation}
L_{dS}=-m_0\frac{ds}{dt}=-m_0\frac{\sqrt{B_{\mu\nu}dx^{\mu}dx^{\nu}}}{dt}=-m_0\sqrt{B_{\mu\nu}\dot{x^{\mu}}\dot{x^{\nu}}}
\end{equation}
where $\dot{x^{\mu}}=\frac{dx^{\mu}}{dt}$ and $B_{\mu\nu}$ is Beltrami metric which can be written as
\begin{equation}
B_{\mu\nu}=\frac{\eta_{\mu\nu}}{\sigma(x)}+\frac{\eta_{\mu\lambda}\eta_{\nu\rho}x^{\lambda}x^{\rho}}{\ell^2\sigma^2(x)},
\quad \eta_{\mu\nu}=dig\{1,-1,-1,-1\}
\end{equation}
where
\begin{equation}
\sigma(x)\equiv 1-\frac{\eta_{\mu\nu}x^{\mu}x^{\nu}}{\ell^2}
\end{equation}
we have Euler-Lagrangian equation
\begin{equation}
\frac{\partial L_{dS}}{\partial x^{i}}-\frac{d}{dt}\frac{\partial L_{dS}}{\partial \dot{x^{i}}}=0
\end{equation}
By employing equations listed above, we can obtain
\begin{equation}
\ddot{x^{i}}=0,\quad \dot{x^{i}}=constant
\end{equation}
This result suggests that the free particle and light signals in BdS spacetime can move along straight line and with constant  coordinate velocities, i.e. inertial motion, and hence the inertial reference frames and inertial observers can be set in BdS spacetime with the inertial Beltrami coordinates $\{x^{\mu}\}$. In order to make sense of inertial motions, the simultaneity should be defined. The two events $A$ and $B$ are defined to be simultaneous if and only if the Beltrami time coordinate $x^0=t$ of the two events are the same, i.e.
\begin{equation}
t(A)=t(B)
\end{equation}
This is called Beltrami simultaneity which defines the laboratory time and the free particles move inertially with respect to it. The $3+1$ decomposition of spacetime can be defined by Beltrami simultaneity
\begin{equation}
ds^2=N^2dt^2-h_{ij}(dx^i+N^idt)(dx^j+N^jdt)
\end{equation}
where $N$ is lapse function, $N^i$ is the i-th component of shift vector and $h_{ij}$ is the induced metric on 3-hypersurface $\Sigma_c$, $i,j=1,2,3$, and
\begin{align}
N&=\left\{\sigma_{\Sigma_c}(x)\left[1-\left(\frac{t}{\ell}\right)^2\right]^{-\frac{1}{2}}\right\}\\
N^i&=tx^i\left[\ell^2-t^2\right]^{-1}\\
h_{ij}&=\delta_{ij}\sigma^{-1}_{\Sigma_c}(x)-\left[\ell\sigma_{\Sigma_c}(x)\right]^{-2}\delta_{ik}\delta_{jg}x^{k}x^{g},(k,g=1,2,3)\\
\sigma_{\Sigma_c}&(x)=1-\left(\frac{t}{\ell}\right)^2+\delta_{ij}x^{i}x^{j}
\end{align}
In addition to Beltrami simultaneity, another simultaneity which is called proper-time($\tau_{\Lambda>0}$) simultaneity with respect to a clock
rest at spatial origin of the Beltrami coordinate system is defined and its cosmological significance is also discussed in Ref.\cite{Guo2,Guo3}, which demonstrate that this simultaneity is closely linked with the cosmological principle if $\tau_{\Lambda>0}$ is regarded as ``cosmic time'', and this predicts that asymptotically the 3-d cosmic
space is slightly closed in order of $\Lambda$ and the further data from WMAP in large scale is needed to support this prediction.

The entropy relations and applications for static Schwarzschild-de Sitter black hole have been investigated in Ref.\cite{xuwei}. In the present paper, we will attempt to construct the first law of thermodynamics and investigate thermodynamical properties  of Schwarzschild-Beltrami-de Sitter black hole with the application of entropy relations. We should note that all the previous research works of the black hole thermodynamics are carried out in the usual non-inertial metric describing  some mixing effects of gravity and inertial force instead of the pure gravity effect arisen from the ``solar mass'' $\mathcal M$, which may make the research results influenced by the extra inertial force. A natural question is that what the black hole thermodynamics will be like if we exclude the effects of inertial force? Or on other hand, we should pay some attention to the black hole thermodynamics generated by pure gravity effects and investigate the possible differences between black holes described by inertial and non-inertial metric. This is key motivation of present work.

\section{Entropy Relations and Entropy Bound}
In this section, we are going to study entropy relations, which can be used to give entropy bound and the first law of thermodynamics such that the Hawking temperature can be read off from the first law. For stationary black holes, we can compute black hole entropy by using the famous Bekenstein-Hawking area entropy \cite{beken} formula

\begin{equation}
S=\frac{\mathcal{A}}{4}\label{4}
\end{equation}
where  $\mathcal{A}$ stands for horizon area of black holes. However, for this S-BdS black hole, the metric tensor is time dependent which means that S-BdS spacetime is not static but dynamical which is different from S-dS black hole. But we can still calculate entropy by formula (\ref{4}) by following two reasons. 1) Sean A.Hayward \textit{et al.} \cite{entropy} proposed that entropy formula (\ref{4}) for black holes in stationary spacetime is also applicable to dynamical black hole and we suppose this relation is still valid for S-BdS black hole. 2) The time dependence of S-BdS metric is attributed to the existence of cosmological constant since we have seen that S-BdS spacetime  would go back to Schwarzschild form which is static when cosmological radius $\ell\rightarrow\infty$, i.e.  cosmological constant $\Lambda\rightarrow 0$. According to current standard model of cosmology, the $\Lambda$-CDM model, includes the cosmological constant which is measured to be on the order of $10^{-52}m^{-2}$ in metric units. Obviously, such a tiny number is not supposed to make some large influences on time independence of the metric of the original Schwarzschild spacetime so that the S-BdS metric can be approximately regarded to be static. Further more, we can also treat the S-BdS black hole as a quasi-static thermodynamical system from the perspective of thermodynamics. Under the consideration as we demonstrate above, we would like to consider time $t$ in metric as a parameter rather than a dynamical variable in our discussion.

\begin{figure}[th]
\centering
\includegraphics[width=3.50in]{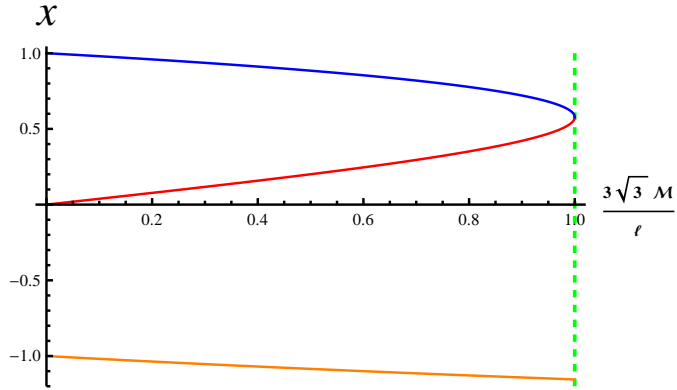}
\caption{The red curve corresponds to $x_1$, the blue curve corresponds to $x_2$ and the orange curve corresponds to $x_3$. All the three curves are plotted in the condition $\ell=1$.\label{fig2}  }
\end{figure}

We can read off  horizon function from the metric (\ref{metric})
\begin{equation}
f(r,t)=1-\frac{r^2}{\ell^2\sigma}-\frac{2\mathcal{M}\sqrt\sigma}{r}\label{3}
\end{equation}
where parameter $\mathcal M$ stands for the mass of the black hole. For horizon radius $r_i$, we have
\begin{equation}
f(r_i,t)=0,(i=1,2,3)
\end{equation}
For solving this equation more easily, we make a simple variable transformation as follows
\begin{align}
x = \frac{r}{\sqrt{\sigma}} ,\quad x^2 = \frac{r^2}{\sigma}\label{2}
\end{align}
Substituting E.q (\ref{2}) into E.q (\ref{3}), $f(r)$ can be rewritten as
\begin{equation}
f(x)=1-\frac{x^2}{\ell^2}-\frac{2\mathcal{M}}{x}
\end{equation}
We denote the roots of the reformed horizon function as $x_i(i=1,2,3)$, which are related to horizon radius $r_i$ by E.q (\ref{2}).
\begin{align}
\begin{dcases}
x_1&=2\ell\sqrt\frac{1}{3} \sin\left({\frac{1}{3}\arcsin{\frac{3\sqrt{3}\mathcal M}{\ell}}}\right)\\
x_2&=2\ell\sqrt\frac{1}{3} \sin\left({\frac{1}{3}\arcsin{\frac{3\sqrt{3}\mathcal M}{\ell}}}+\frac{2\pi}{3}\right)\\
x_3&=2\ell\sqrt\frac{1}{3} \sin\left({\frac{1}{3}\arcsin{\frac{3\sqrt{3}\mathcal M}{\ell}}}-\frac{2\pi}{3}\right)\label{roots}
\end{dcases}
\end{align}
These three roots $x_1,x_2,x_3$ above are just the event horizon, cosmological horizon and virtual horizon(since $x_3<0$) in S-dS spacetime. We plot a diagram in Fig.(\ref{fig2}) to compare the three roots at fixed $\ell$. From the diagram, it is obviously to see that $x_3$ is always keeping negative but it has greater absolute value than both $x_1$ and $x_2$ while $x_2$ is greater than $x_1$ until $3\sqrt 3\mathcal M\ell^{-1}$ approaches one. This fact could be summarized as
\begin{equation}
0\leq x_1\leq x_2\leq |x_3|\label{xroot}
\end{equation}
This result will be used in the following discussion.

\begin{figure}[th]
\centering
\includegraphics[width=2.1in]{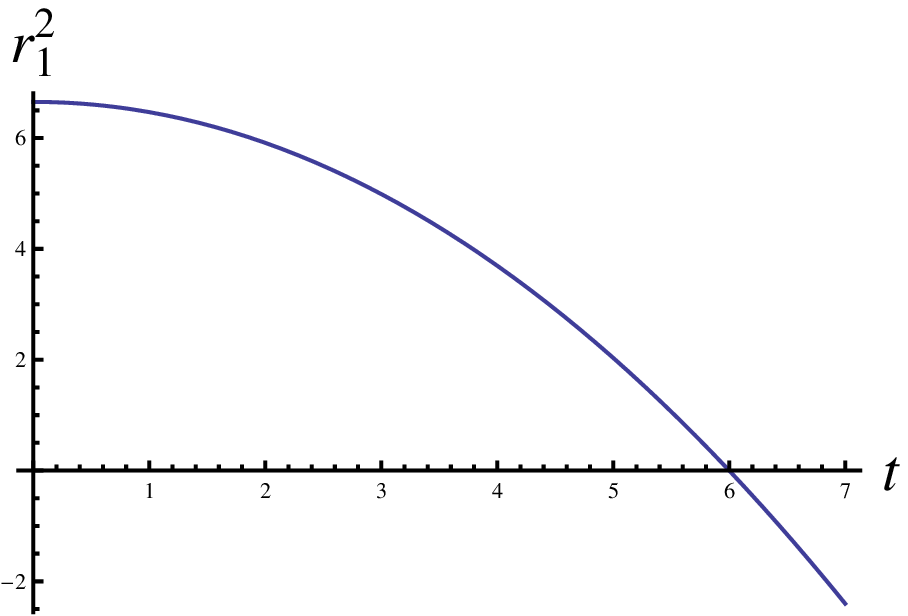}~~
\includegraphics[width=2.1in]{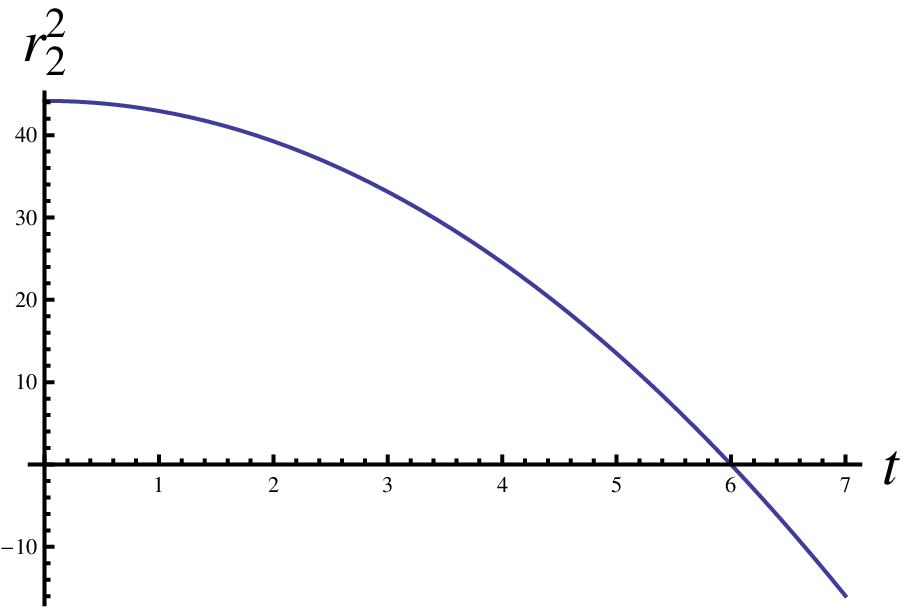}~~
\includegraphics[width=2.1in]{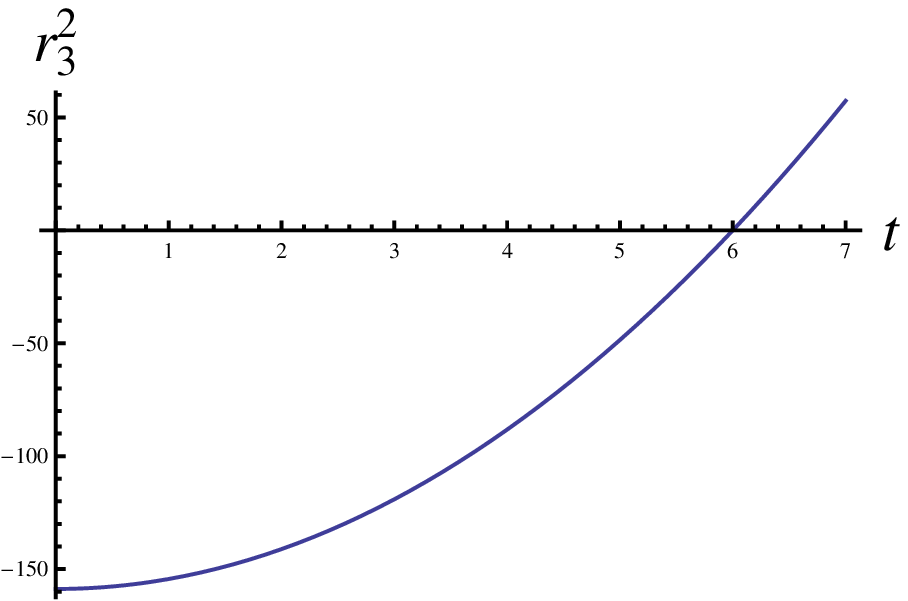}
\caption{The $r_i^2-t$ diagram. Here we take $\ell=6$, $\mathcal M=1$ and obviously, black hole mass and cosmological radius are restricted to $3\sqrt{3}\mathcal M\leq \ell$.\label{fig1}}
\end{figure}

By employing E.q (\ref{2}), it is directly to obtain horizon radius $r_i$
\begin{align}
\begin{dcases}
r_1^2&=\frac{4(\ell^2-t^2)\sin^2\left(\frac{1}{3}\arcsin(\frac{3\sqrt{3}\mathcal{M}}{\ell})\right)}{1+2\cos\left({\frac{2}{3}}\arcsin({\frac{3\sqrt{3}\mathcal M}{\ell}})\right)}\label{r1}\\
r_2^2&=\frac{4(\ell^2-t^2)\cos^2\left(\frac{1}{3}\arcsin(\frac{3\sqrt{3}\mathcal{M}}{\ell})+\frac{\pi}{6}\right)}{3-4\cos^2\left({\frac{1}{3}}\arcsin({\frac{3\sqrt{3}\mathcal M}{\ell}})+\frac{\pi}{6}\right)}\\
r_3^2&=\frac{4(\ell^2-t^2)\cos^2\left(-\frac{1}{3}\arcsin(\frac{3\sqrt{3}\mathcal{M}}{\ell})+\frac{\pi}{6}\right)}{3-4\cos^2\left(-{\frac{1}{3}}\arcsin({\frac{3\sqrt{3}\mathcal M}{\ell}})+\frac{\pi}{6}\right)}
\end{dcases}
\end{align}
For horizon radius $r_1$, we have
\begin{equation}
\lim_{\ell\rightarrow\infty}r_1^2=4\mathcal M^2
\end{equation}
which is nothing but the squared event horizon radius of the Schwarzschild black hole. In order to show the relationship between horizon radius $r_i$ and time parameter $t$, we plot $r_i^2-t$ diagram in Fig.(\ref{fig1}). From the diagrams we can see that for horizon radius $r_1$ and $r_2$, the time $t$ must be restricted to $t\leq\ell$ to keep squared horizon radius positive, while for $r_3$, $t\geq\ell$ is the sufficient condition to make $r_3^2$ remain positive. Comparing the the diagram for $r_1^2$ with diagram for $r_2^2$, we find that radius $r_2$ is always greater than radius $r_1$ except at the point $t=\ell$ where the two horizon radius both reduce to zero.

Although horizon radius is dependent of time parameter $t$, the area of the three horizons remains constant as we have horizon area
\begin{equation}
\begin{split}
\mathcal A_{i}&=\int\int\sqrt{g_{\theta\theta}g_{\phi\phi}}d\theta d\phi, (i=1,2,3)\\
&=4\pi\frac{r_i^2}{\sigma}=4\pi x_i^2
\end{split}
\end{equation}
entropy on the horizons is computed by formula (\ref{4})
\begin{equation}
S_i=\frac{\mathcal A_i}{4}=\pi x_i^2\label{entropy}
\end{equation}

Entropy relations of black holes have been studied widely \cite{ppr,xuwei,wj,root,Gao1,Tian1,Liu2016,LiuHang} and we will revisit these relations and show that the S-BdS black hole owns the same entropy relations as S-dS black hole. With the help of E.q(\ref{roots}) and (\ref{entropy}), entropy relations can be calculated directly or by the use of Vieta theorem. For example, the mass-dependence entropy product is
\begin{equation}
S_1 S_3 S_3=4\ell^4 \mathcal M^2\pi^3\label{product}
\end{equation}
and the mass-independence relations \cite{wj}
\begin{align}
&S_1+S_2+S_3=2\ell^2\pi\label{sum}\\
&S_1S_2+S_1S_3+S_2S_3=\ell^4\pi^2\label{pproduct}
\end{align}
Obviously, all these three relations above are independent of time. Note that $0\leq x_1\leq x_2\leq|x_c|$ (E.q \ref{xroot}), we get
\begin{equation}
S_1\leq S_2\leq S_3=\pi x_3^2\leq \frac{4}{3}\ell^2
\end{equation}
From the root relation
\begin{equation}
x_1+x_2+x_3=0\label{5}
\end{equation}
which can be obtained by straightforward  calculation or  Vieta theorem we can easily get the following relation
\cite{xuwei}

\begin{equation}
S_1+S_2+\sqrt{S_1S_2}=\frac{1}{2}(S_1+S_2+S_3)=\pi\ell^2
\end{equation}
which yields
\begin{gather}
0\leq3S_1\leq (S_1+S_2+\sqrt{S_1S_2})=\ell^2\pi\leq 3S_2
\intertext{and}
S_2\leq(S_1+S_2+\sqrt{S_1ES_2})=\ell^2\pi\\
S_2\leq S_3\leq\frac{4}{3}\ell^2\pi
\end{gather}
Considering the combination of equations above, we can get the entropy bound on the horizons
\begin{equation}
S_1\in[0,\frac{\pi}{3}\ell^2],\quad S_2\in[\frac{\pi}{3}\ell^2,\pi\ell^2],\quad S_3\in[\pi\ell^2,\frac{4\pi}{3}\ell^2]\label{16}
\end{equation}
which are all characterized by geometrical parameter $\ell$.

Since $\frac{\partial }{\partial t}$ is no longer a killing vector filed in S-BdS spacetime implying that the spacetime is dynamic which leads to the Hawking temperature \cite{hk} can not be obtained by compute the surface gravity defined on killing horizon for stationary black hole in this spacetime, such that we can not get temperature by usual formula

\begin{equation}
T_h=\frac{k}{2\pi}
\end{equation}
where $k$ represents the surface gravity on killing horizons.
On the other hand, we still believe that this non-stationary black hole can radiate energy as it's stationary counterpart does. Actually, people have established the concept of weakly isolated horizon and the corresponding zeroth and first law of thermodynamics for studying non-stationary black hole thermodynamics but we do not intend to adopt that concept to do calculation due to the S-BdS black hole is not rigorously regarded  as a dynamical black hole in our discussion  and we would like to work out the first law and temperature of S-BdS black hole by turning to entropy relations analysis.

Consider carrying out a perturbation to entropy relations (\ref{product},\ref{sum},\ref{pproduct})
\begin{align}
&\delta S_1+\delta S_2+\delta S_3=-\frac{6\pi}{\Lambda^2}\delta\Lambda\\
&S_1\delta S_3+S_2\delta S_1+S_1\delta S_3+S_3\delta S_1+S_2\delta S_3+S_3\delta S_2=-\frac{18\pi^2}{\Lambda^2}\delta \Lambda\\
&S_1S_2\delta S_3+S_1S_3\delta S_2+S_2S_3\delta S_1=-\frac{72\mathcal M^2\pi^3}{\Lambda^3}\delta\Lambda+\frac{72\mathcal M\pi^3}{\Lambda^2}\delta\mathcal M
\end{align}
where the cosmological radius $\ell$ has been replaced with cosmological constant $\Lambda$ by relation $\Lambda=\frac{3}{\ell^2}$ and we have treated cosmological constant $\Lambda$ as a thermodynamical variable. \cite{dk,bp,mc,bpd,na,Liu0}. These lead to
\begin{align}
\delta\mathcal M&=\frac{(S_1-S_2)(S_1-S_3)\Lambda^3}{72\mathcal M\pi^3\Lambda}\delta S_1+\frac{12\mathcal M^2\pi^2+S_1(S_1\Lambda-3\pi)}{12\mathcal M\pi^2\Lambda}\delta\Lambda\label{6}\\
\delta\mathcal M&=\frac{(S_1-S_2)(S_3-S_2)\Lambda^3}{72\mathcal{M}\pi^3\Lambda}\delta S_2+\frac{12\mathcal M^2\pi^2+S_2(S_2\Lambda-3\pi)}{12\mathcal M\pi^2\Lambda}\delta\Lambda\label{7}\\
\delta\mathcal M&=\frac{(S_1-S_3)(S_2-S_3)\Lambda^3}{72\mathcal M\pi^3\Lambda}\delta S_3+\frac{12\mathcal M^2\pi^2+S_3(S_3\Lambda-3\pi)}{12\mathcal M\pi^2\Lambda}\delta\Lambda\label{8}
\end{align}

It is well known that a positive cosmology  constant $\Lambda$ is related to a negative pressure $P$ which is responsible for the acceleration expansion of the universe by relation

\begin{equation}
P=-\frac{\Lambda}{8\pi}\label{P}
\end{equation}
Substituting E.q (\ref{P}) into E.q (\ref{6},\ref{7},\ref{8}), we have the First Law
\begin{gather}
\begin{dcases}
\delta\mathcal M=T_1\delta S_1+V_1\delta P\label{L1}\\
\delta\mathcal M=T_2\delta S_2+V_2\delta P\\
\delta\mathcal M=T_3\delta S_3+V_3\delta P
\end{dcases}
\end{gather}
where
\begin{gather}
\begin{dcases}
T_1=\frac{(S_1-S_2)(S_1-S_3)\Lambda^3}{72\mathcal M\pi^3\Lambda}\\
T_2=\frac{(S_1-S_2)(S_3-S_2)\Lambda^3}{72\mathcal M\pi^3\Lambda}\\
T_3=\frac{(S_1-S_3)(S_2-S_3)\Lambda^3}{72\mathcal M\pi^3\Lambda}
\end{dcases}
\end{gather}
and
\begin{gather}
\begin{dcases}
V_1=\frac{-24\mathcal M^2\pi^2+2S_1(3\pi-S_1\Lambda)}{3\mathcal M\pi\Lambda}\\
V_2=\frac{-24\mathcal M^2\pi^2+2S_2(3\pi-S_2\Lambda)}{3\mathcal M\pi\Lambda}\\
V_3=\frac{-24\mathcal M^2\pi^2+2S_3(3\pi-S_3\Lambda)}{3\mathcal M\pi\Lambda}
\end{dcases}
\end{gather}
We just simply regard E.qs (\ref{L1}) as the first law of thermodynamics on horizon $r_1,r_2$ and $r_3$ of S-BdS black hole respectively, and obviously, the parameter $\mathcal M$ should be interpreted as enthalpy rather than mass in the first law. It is easy to check that the temperature and thermodynamical volume $V_i$ we obtain above can be simply written as
\begin{align}
&T_i=\frac{f'(x)}{4\pi}\bigg |_{x=x_i}=\frac{1}{4\pi}\frac{\ell^2-3x_i^2}{\ell^2x_i}\label{10}\\
&V_i=\frac{4}{3}\pi x_i^3
\end{align}
where the $f'(x)$ represents the derivative of function $f(x)$ with respect to $x$. We substitute $\sigma=1-\frac{t^2-r^2}{\ell^2}$ into E.q (\ref{10}) to obtain
\begin{gather}
T_i=\frac{1}{4\pi}\frac{\ell^2-3x_i^2}{\ell^2x_i}=\frac{\ell^2-2r_i^2-t^2}{4\ell^2\pi r_i\sqrt{\frac{\ell^2+r_i^2-t^2}{\ell^2}}}\label{12}
\end{gather}
One can check that the temperature function goes back to the form in the case of Schwarzschild black hole when $\ell\rightarrow \infty$

\begin{equation}
\lim_{\ell\rightarrow \infty}\frac{\ell^2-2r_i^2-t^2}{4\ell^2\pi r_i\sqrt{\frac{\ell^2+r_i^2-t^2}{\ell^2}}}=\frac{1}{4\pi r_i}\label{11}
\end{equation}
\\
This result is just the Hawking temperature for event horizon of Schwarzschild black hole. Further more, for the heat capacity $C_{h}=\frac{\partial M}{\partial T}$, it just has the same behavior when $\ell\rightarrow \infty$ like temperature function does. So it is reasonable to take (\ref{11}) as temperature for corresponding horizons of  Schwarzschild-Beltrami-de Sitter black hole. We also have Samrr relations
\begin{align}
\begin{dcases}
\mathcal M&=2(T_1S_1-V_1P)\label{Smarr}\\
\mathcal M&=2(T_2S_2-V_2P)\\
\mathcal M&=2(T_3S_3-V_3P)
\end{dcases}
\end{align}
Finally, we obtain the entropy bound (\ref{16}), the first law of thermodynamics (\ref{L1}) and Smarr relations (\ref{Smarr}) for the three horizons(horizon $r_1$, $r_2$ and $r_3$) of Schwarzschild-Beltrami-de Sitter black hole and all these results are derived by the entropy relation analysis.

\section{Thermodynamical Analysis}
In this section, we focus on the thermodynamical analysis on horizon $r_1$ for the reason that behavior of this horizon is similar to the event horizon of Schwarzschild black hole. We rewrite the first law on horizon $r_1$ as
\begin{equation}
\delta M=T_h\delta S+V_h\delta P
\end{equation}
where $T_h$ and $V_h$ stand for the temperature and the thermodynamical volume for horizon $r_1$, and for convenience, we also denote $r_1$ as $r_h$. To have a better understanding of the thermodynamics of S-BdS black hole, it is necessary to investigate the behavior of the thermodynamical quantities, such as Gibbs free energy, heat capacity and temperature. And by comparing these thermodynamical quantities of S-BdS black hole with those corresponding quantities of S-dS black hole, we can clearly illustrate the differences between the two black holes which  reflects the the property of the thermodynamics of black holes described by inertial coordinates eliminating the effects of inertial force. We should notice that since we focus on the thermodynamical analysis on horizon $r_1$ which is limited by E.q(\ref{r1})and has the bound
\begin{equation}
0<r_h\leq\left(\frac{1}{2}(\ell^2-t^2)\right)^{\frac{1}{2}}\label{lr1}
\end{equation}
and similarly, for S-dS black hole, the event horizon radius $r_h$ is restricted to
\begin{equation}
0<r_h\leq \frac{\ell}{\sqrt 3}\label{lrh}
\end{equation}
As we would like to take $\ell=4$ and $t=0$ in the following discussion, which makes the $r_h$  restricted to $0<r_h\leq2\sqrt 2$ (for S-BdS black hole) and $0<r_h\leq \frac{4}{\sqrt 3}$(for S-dS black hole).

\subsection{Gibbs Free Energy}
In the thermodynamics of black holes, Gibbs free energy diagram plotted at the fixed pressure $P$ in the $G-T$ plane is usually used to investigate the smll/large black hole phase transition indicated by the characteristic swallowtail behavior of $G-T$ curve, or phase transition between black holes and radiation which usually occurs in the case of uncharged black holes. In fact, the first law (\ref{L1}) we have obtained is exactly identical with the first law of AdS black holes with the exception that the pressure $P=-\frac{\Lambda}{8\pi}$ is negative now. This implies that we can understand the Gibbs free energy of de Sitter black holes as a constitution of the AdS case to the region of negative pressure \cite{dS}.

\begin{figure}[!htb]
\centering
\includegraphics[width=3in]{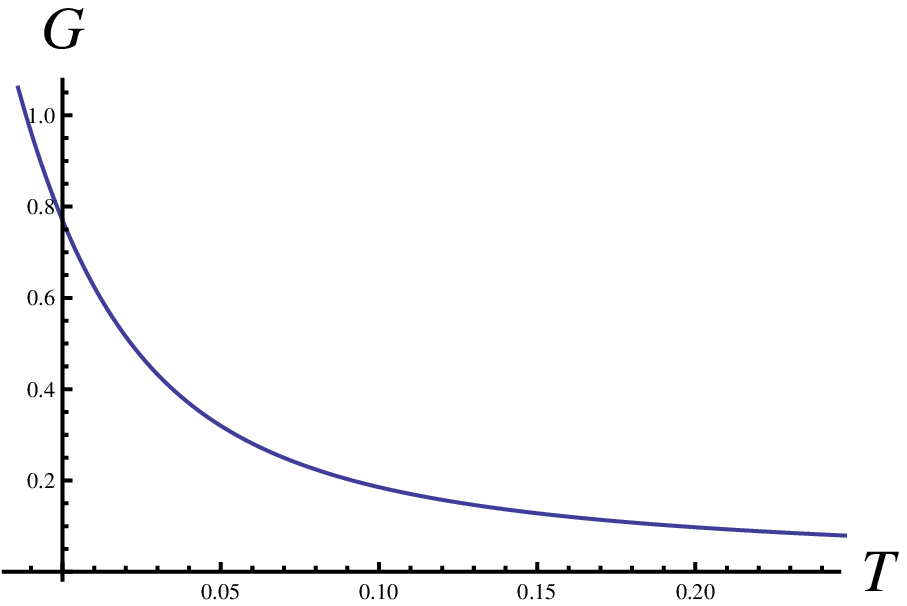}~~
\includegraphics[width=3in]{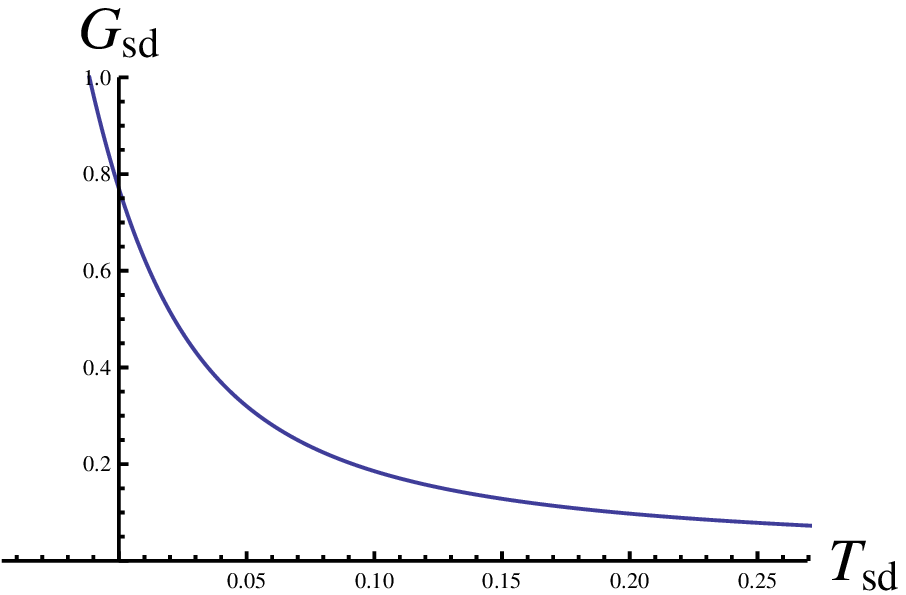}
\caption{The $G-T$ diagram, the left plot corresponds to S-BdS black hole and the right plot corresponds to S-dS black hole, where the $T_{sd}$ represents the Hawking temperature on event horizon of S-dS black hole. Here we take $\ell=4$ and $t=0$.\label{fig3}}
\end{figure}

\begin{figure}[!htb]
\centering
\includegraphics[width=3.5in]{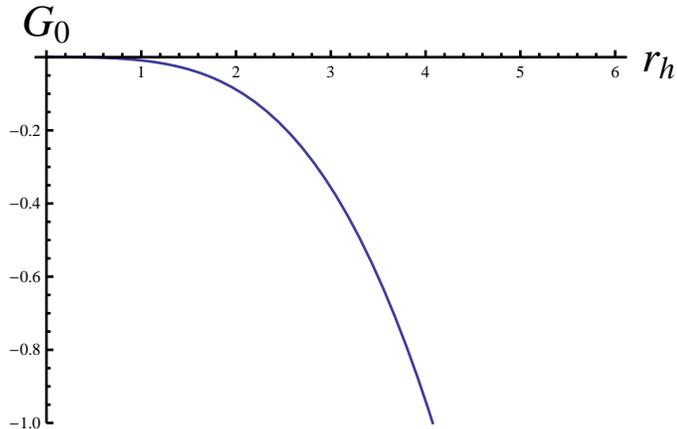}
\caption{The $G_0-r_h$ diagram, the plot shows a negative definite difference between the Gibbs free energy of S-BdS and S-dS black hole with the condition that $t=0$ and $\ell=4$.\label{fig4}}
\end{figure}

For S-BdS  black hole, the Gibbs free energy has the form
\begin{equation}
\begin{split}
G&=\mathcal M-T_hS=\frac{r_h(\ell^2+2r_h^2-t^2)}{4\ell^2}\left(\frac{\ell^2+r_h^2-t^2}{\ell^2}\right)^{-\frac{3}{2}}
\end{split}
\end{equation}
and the Gibbs free energy for S-dS black hole
\begin{equation}
G_{sd}=\frac{r_h(\ell^2+r_h^2)}{4\ell^2}
\end{equation}
We plot Gibbs energy as a function of temperature $T$ in Fig.(\ref{fig3}). From the diagram, we can see that the Gibbs free energy $G$ of S-BdS black hole has almost the same behavior as $G_{sd}$: both of them are monotone decrease and positive, no criticality point and no phase transition characteristic caused by the negative pressure $P$.
In order to find out the subtle distinctions between $G$ and $G_{sd}$, we study  $\Delta G=G-G_{sd}$
\begin{equation}
\begin{split}
\Delta G&=G-G_{sd}=-\frac{1}{4\ell^2}r_h\left\{r_h^2(r_h^2-t^2)\sqrt{\frac{\ell^2+r_h^2-t^2}{\ell^2}}\right.\\
&+\left.\left(\sqrt{\frac{\ell^2+r_h^2-t^2}{\ell^2}}-1\right)\{\ell^2(2r_h^2-t^2)+\ell^4\}\right\}\left(\frac{\ell^2+r_h^2-t^2}{\ell^2}\right)^{-\frac{3}{2}}
\end{split}
\end{equation}
We would like to consider the case that $t=0$ and $\ell=4$ which leads to
\begin{equation}
\Delta G_{0}=-\frac{r_h^3}{64}+r_h\left(\frac{2}{\sqrt{16+r_h^2}}-\frac{16}{(16+r_h^2)^{\frac{3}{2}}}-\frac{1}{4}\right)
\end{equation}
and we plot the $\Delta G_0-r_h$ diagram in Fig.(\ref{fig4}) which shows that the value of $G-G_sd$ is always negative which means that the Gibbs free energy($G$) of S-BdS black hole is smaller than Gibbs free energy($G_{sd}$) of S-dS black hole at the same horizon radius $r_h$ and this fact implies that the S-BdS black hole is more stable than S-dS black hole.
\subsection{Heat Capacity}
The heat capacity at constant pressure $C_P$ of black holes can be used to distinguish  the black hole is stable or unstable, i.e., positive heat capacity corresponds to stable black holes and the negative heat capacity corresponds to unstable black holes. What's more, the divergence of the heat capacity in the $G_P-r_h$ is usually used as a characteristic of phase transition.

For S-BdS black hole, we have the heat capacity at constant pressure
\begin{equation}
C_P=\frac{\partial \mathcal M}{\partial r_h}\left(\frac{\partial T_h}{\partial r_h}\right)^{-1}=
\frac{2\pi\ell^2(2r_h^2+t^2-\ell^2)}{(\ell^2+r_h^2-t^2)(\ell^2+4r_h^2-t^2)}\label{heat1}
\end{equation}
for S-dS black hole, the heat capacity is
\begin{equation}
C_P=-\frac{2\pi r_h^2(\ell^2-3r_h^2)}{\ell^2+3r_h^2}\label{heat2}
\end{equation}
Particularly, for both S-BdS black hole and S-dS black hole, we solve equation in the condition $t=0$ and $\ell=4$
\begin{equation}
C_P=0\label{H}
\end{equation}
the root of E.q(\ref{H}) is $r_h=2\sqrt 2$ for S-BdS black and $r_h=\frac{4}{\sqrt 3}$ for S-dS black hole,  coincides with the two black hole's upper bound of the horizon radius(\ref{lr1},\ref{lrh}), respectively.

\begin{figure}[!htb]
\centering
\includegraphics[width=3.5in]{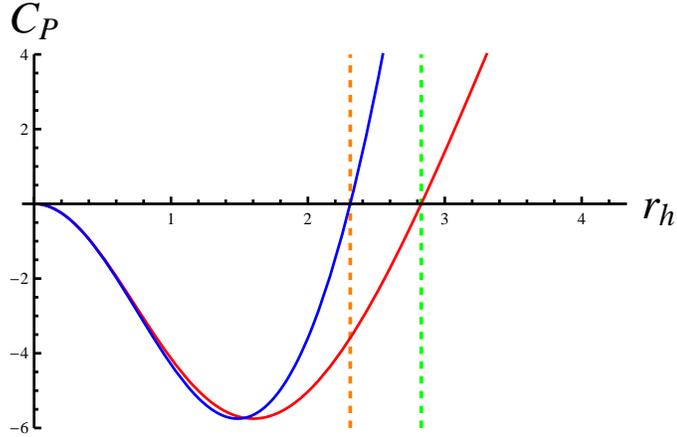}
\caption{The $C_P-r_h$ diagram, the red curve corresponds to S-BdS black hole and the blue curve corresponds to S-dS black hole. The abscissa point $r_h=2\sqrt 2$(green dashed straight line) and $r_h=\frac{4}{\sqrt 3}$(orange dashed straight line) correspond to the zero value of the heat capacity for the two black holes, respectively.  Here we still take $t=0$ and $\ell=4$.\label{fig5}}
\end{figure}

We plot $C_P-r_h$ diagram in Fig.(\ref{fig5}) from which we could see that the behaviors of heat capacity of the two black holes are similar, as both of them are smooth and no divergence point indicating that no phase transition occurs as expected. On the other hand, under the consideration of E.q(\ref{lr1}, \ref{lrh}), both  the S-dS black hole and S-BdS black hole would not undergo a transition from unstable state to the stable state like AdS black hole does, i.e., the heat capacity remains  negative when  horizon radius of the two black holes within bounds as we show in the diagram.

\begin{figure}[!htb]
\centering
\includegraphics[width=3.5in]{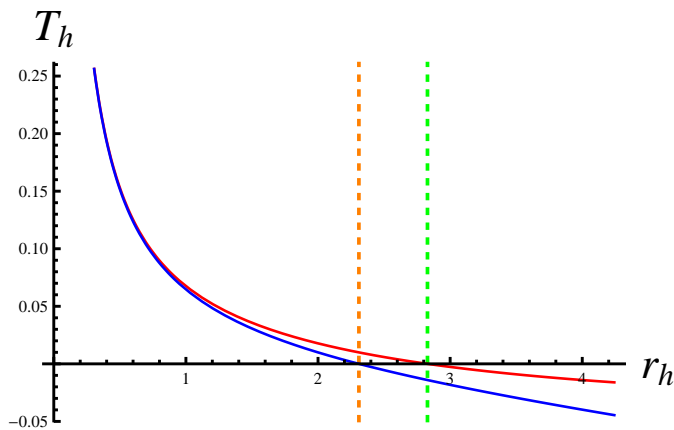}
\caption{The $T_h-r_h$ diagram, the red curve corresponds to S-BdS black hole and the blue curve corresponds to S-dS black hole. The  point $r_h=2\sqrt 2$(green dashed straight line) and $r_h=\frac{4}{\sqrt 3}$(orange dashed straight line) at  abscissa correspond to the zero value of the temperature for the two black hole, respectively.  Here we still take $t=0$ and $\ell=4$.\label{fig6}}
\end{figure}

The behavior of heat capacity can also be reflected by temperature function since the derivative of temperature with respect to horizon radius $r_h$ plays the role of denominator in the heat capacity (\ref{heat1}, \ref{heat2}). Just as do to heat capacity, we have following equations for S-BdS black hole in the condition $t=0, \ell=4$
\begin{gather}
T_h(r_h)=\frac{\ell^2-2r_h^2-t^2}{4\ell^2\pi r_h\sqrt{\frac{\ell^2+r_h^2-t^2}{\ell^2}}}=0\\
r_h=2\sqrt{2}
\end{gather}
and for S-dS black hole in the same condition
\begin{gather}
T_h=\frac{1}{4\pi}\frac{\ell^2-3r_h^2}{\ell^2r_h}=0\\
r_h=\frac{4}{\sqrt 3}
\end{gather}
the two roots still coincide with the upper bound of the horizon radius of S-BdS and S-dS black hole respectively. Note that the upper bound of the horizon radius corresponds to the extreme condition
\begin{equation}
\mathcal{M}=\frac{\ell}{3\sqrt 3}
\end{equation}
which leads to the coincidence of horizon $r_1$ and $r_2$. This result is consistent with the characteristic of extreme black holes, e.g., the R-N black holes which have two horizons $r_{\pm}=\mathcal{M}\pm\sqrt {\mathcal M^2-Q^2}$, when satisfying the extreme condition $\mathcal M=Q$, the two horizons $r_{\pm}$ coincide with each other as $r_+=r_-=\mathcal M$ with Hawking temperature  $T_h=0$. In this sense, the cosmological radius plays a role of electric charge in S-BdS/S-dS black holes. We plot the $T_h-r_h$ diagram in Fig.(\ref{fig6}). As expected, the diagram shows that the temperatures of both black holes are positive when their respective horizon radius within bounds and are monotone decreasing with the increase of $r_h$, and at the same $r_h$, the temperature of S-BdS black hole(red curve) is a little higher than the temperature of S-dS black hole(blue curve).

Finally, we have analyzed the thermodynamical quantities of  Schwarzschild-Beltrami-de Sitter black hole and  Schwarzschild-de Sitter black hole with the finding that the two black holes appear to have similar thermodynamical properties provided that the entropy formula (\ref{4}) is viable for S-BdS black holes.

\section{Conclusions and Discussions}
In this paper, we have investigated the thermodynamical properties of Schwarzschild-Beltrami-de Sitter black hole which is introduced by Mu-Lin Yan \textit{et al.} in 2013 by introducing inertial Beltrami coordinates to traditional non-inertial Schwarzschild-de Sitter metric which is the exact static spherical symmetry solution of Einstein equation with a positive cosmological constant $\Lambda$. Based on this new metric, we have computed entropies on all horizons and obtained entropy bound  which only depends on the geometrical parameter $\ell$ for each horizon. Hawking temperatures and the first law of thermodynamics are obtained  by considering a perturbation to entropy relations due to that the spacetime described by this inertial coordinates is no longer a stationary spacetime in which surface gravity related to Hawking temperature is defined well on killing horizon. The Smarr relations are also derived. To have a deeper understanding of the S-BdS black hole, we have studied the  Gibbs free energy, heat capacity and temperature function which behave highly similar to S-dS black hole, no any phase transition characteristics are found. However, we find that at the same horizon radius, the Gibbs free energy of S-BdS black hole is smaller than that of S-dS black hole which implies that the S-BdS black hole is more stable than S-dS black hole. When extreme condition is satisfied, we find that the cosmological radius $\ell$ can play a role of electric charge as the temperatures of both black holes would drop to zero, which is the characteristic of the extreme black holes.

As the results we have concluded in our discussions  show, the S-BdS black hole described by  inertial Beltrami coordinates which is introduced to exclude the inertial force included in the S-dS metric, seems to have similar thermodynamical properties to S-dS black hole, which means that the inertial force may not have great influences on black hole thermodynamics. The authors in Ref.\cite{Guo1} argued that the temperature defined by the inertial observers in BdS spacetime is zero and the entropy on horizon also should be zero although there exists a horizon and it is not needed
to search for the statistical origin of the entropy. However, one may note that when S-BdS black hole mass $\mathcal M=0$, the temperature on cosmological horizon $x_2$ and virtual horizon $x_3$ will not vanish by using E.q (\ref{10}) and (\ref{roots}), and the entropy on corresponding horizon does not vanish, either. On the one hand, this problem comes from the entropy formula (\ref{4}) we use in this paper. In fact, the Hawking-Bekenstein area entropy is only discussed in the spacetime described by non-inertial metric, for the inertial coordinates, such as Beltrami coordinates, the entropy formula should probably be corrected. On the other hand, the black hole thermodynamics itself is not perfect, for example, the hawking  temperature of extreme RN black hole is zero while there exists horizon which leads to non-zero entropy on horizon by using the usual entropy formula $S=\mathcal{A}/4$, but this result contradicts our usual understanding of thermodynamics i.e. zero temperature corresponds to zero entropy. And in the Schwarzschild black hole case, the temperature of which is expressed as $T=1/8\pi \mathcal M$, one can find that when $\mathcal{M}\rightarrow0$, the temperature $T\rightarrow\infty$ although the spactime would go back to  Minkowski flat spactime which is supposed to have zero temperature for the inertial observer. These facts suggest that the black hole thermodynamics should probably be developed further. In this sense, our present work may shed some light on the deeper understanding of S-BdS black hole thermodynamics which is worthy to be explored further and this is left to be a future work.

\section*{Acknowledgements}
This project is partially supported by NSFC.
\bibliography{dbsd}
\end{document}